\documentclass[ALICE,manyauthors]{cernphprep}
\usepackage[comma,square,numbers,sort&compress]{natbib}
\usepackage{hyperref}
\usepackage{lineno}
\usepackage{color}
\usepackage{caption}
\usepackage{multirow}

\usepackage{graphics}
\usepackage{epstopdf}
\usepackage{amssymb}
\usepackage{xspace}
\usepackage{amsmath}
\usepackage{upgreek}
\usepackage{hyperref}
\usepackage[T1]{fontenc}

\usepackage{doi}
\usepackage{orcidlink}

\newcommand{\snn}{$\sqrt{s_{\mathrm{NN}}}~$}

\newcommand{\pt}{$p_{\mathrm{T}}~$}

\newcommand{\ptt}{\ensuremath{p_{\mathrm{T}}^{\rm trig}}\xspace}

\newcommand{\pp}           {pp\xspace}

\newcommand{\PbPb}         {\mbox{Pb--Pb}\xspace}

\newcommand{\pPb}          {\mbox{p--Pb}\xspace}

\newcommand{\GeVc}         {Ge\kern-.1emV/$c$\xspace}
\newcommand{\MeVc}         {Me\kern-.1emV/$c$\xspace}
\newcommand{\TeV}          {Te\kern-.1emV\xspace}
\newcommand{\GeV}          {Ge\kern-.1emV\xspace}
\newcommand{\MeV}          {Me\kern-.1emV\xspace}
\newcommand{\GeVmass}      {Ge\kern-.2emV/$c^2$\xspace}
\newcommand{\MeVmass}      {Me\kern-.2emV/$c^2$\xspace}

\newcommand{\SPD}          {\rm{SPD}\xspace}

\newcommand{\VZERO}        {\rm{V0}\xspace}
\newcommand{\VZEROA}       {\rm{V0A}\xspace}
\newcommand{\VZEROC}       {\rm{V0C}\xspace}

\usepackage{float} 

\begin{document}%
\begin{titlepage}

\PHyear{2022}
\PHnumber{059}      
\PHdate{22 March}  
  

\title{Study of charged particle production at high $p_{\rm T}$ using event topology in \pp, \pPb and \PbPb collisions at $\sqrt{\mathbf{s_{\rm NN}}}$ = 5.02\,\TeV}
\ShortTitle{High-$p_{\rm T}$ charged-particle production using event topology in \pp, \pPb and \PbPb}
\Collaboration{ALICE Collaboration\thanks{See Appendix~\ref{app:collab} for the list of collaboration members}}
\ShortAuthor{ALICE Collaboration} 
\begin{abstract}
This letter reports measurements which characterize the underlying event associated with hard scatterings at mid-pseudorapidity ($|\eta|<0.8$) in \pp, \pPb and \PbPb collisions at centre-of-mass energy per nucleon pair, $\sqrt{s_{\rm NN}}=5.02$\,\TeV. The measurements are performed with ALICE at the LHC. Different multiplicity classes are defined based on the event activity measured at forward rapidities. The hard scatterings are identified by the leading particle defined as the charged particle with the largest transverse momentum ($p_{\rm T}$) in the collision and having $8$ $< p_{\rm T}<15$\,\GeVc. The $p_{\rm T}$ spectra of associated particles ($0.5$ $\leq p_{\rm T}<6$\,\GeVc) are measured in different azimuthal regions defined with respect to the leading particle direction: toward, transverse, and away.  The associated charged particle yields in the transverse region are subtracted from those of the away and toward regions.  The remaining jet-like yields are reported as a function of the multiplicity measured in the transverse region. The measurements show a suppression of the jet-like yield in the away region and an enhancement of high-$p_{\rm T}$ associated particles in the toward region in central \PbPb collisions, as compared to minimum-bias \pp collisions. These observations are consistent with previous measurements that used two-particle correlations, and with an interpretation in terms of parton energy loss in a high-density quark gluon plasma. These yield modifications vanish in peripheral \PbPb collisions and are not observed in either high-multiplicity \pp or \pPb collisions.

\end{abstract}
\end{titlepage}
\setcounter{page}{2}
\section{Introduction}\label{sec:intro}

\begin{figure}[ht!]
\centering
\includegraphics[width=12pc]{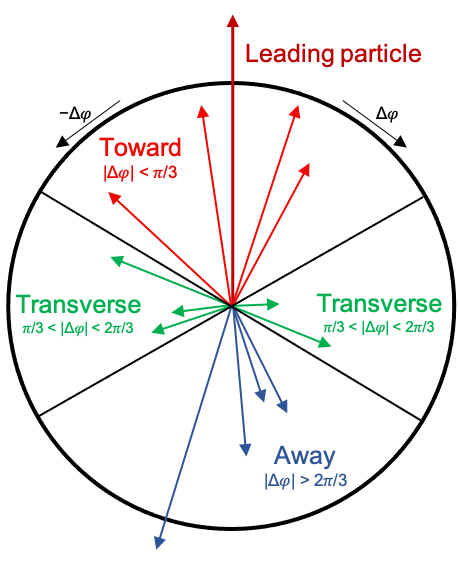}
\caption{(colour online) Illustration of toward, away and transverse regions with respect to the leading particle in a collision.
}
\label{Regions}
\end{figure}

In proton-proton (\pp) collisions, jets, originating from partonic scatterings with large momentum transfer, are accompanied by particles produced by initial- and final-state radiation (ISR and FSR, respectively), as well as, by a plethora of other mechanisms. These include proton break-up, and, in a scenario incorporating multi-parton interactions (MPI)~\cite{Sjostrand:1987su,Bartalini:2017jkk}, several semi-hard parton-parton scatterings in a single \pp collision. These jet-accompanying particles experimentally make up the underlying event (UE) and are commonly studied via azimuthal separations from the jets to minimise the influence of hard scatterings. The present study follows the strategy originally introduced by the CDF collaboration~\cite{Affolder:2001xt}. First, the leading charged particle in the event is found, i.e., the charged particle with the highest transverse momentum in the collision (\ptt). Secondly, the associated particles (\pt<\ptt) are measured in three topological regions depending on their azimuthal angle relative to the leading particle, $|\Delta\varphi| =|\varphi^{\mathrm{assoc}}-\varphi^{\mathrm{trig}}|$, see Fig.~\ref{Regions}.

The toward region contains the primary jet  within the acceptance of the detector, while the away region contains the back-scattered particles of the recoil jet~\cite{Adam:2019xpp}. In contrast, the transverse region is dominated by the underlying-event dynamics, but it also includes contributions from ISR and FSR~\cite{Buttar:2005gdq}. 

The measurements performed at RHIC and LHC in small systems (\pp, p--A, and d--A collisions) have shown for high particle multiplicities similar phenomena as were originally observed only in A--A collisions and have been attributed there to the formation of the strongly interacting quark gluon plasma~\cite{STAR:2005gfr,PHENIX:2004vcz}, namely, long range angular correlations and collectivity~\cite{Busza:2018rrf}. The origin of these effects in small systems is still an open question; on one hand, hydrodynamical calculations describe some aspects of the data~\cite{Nagle:2018nvi}; on the other hand,  mechanisms like colour reconnection~\cite{OrtizVelasquez:2013ofg}, rope hadronisation~\cite{Bierlich:2014xba}, and string shoving~\cite{Bierlich:2020naj} can produce collective-like effects in Monte Carlo event generators such as \textsc{PYTHIA}~8~\cite{Sjostrand:2014zea}. 
Thus, investigating \pp collisions as a function of the charged particle multiplicity has become ever more pertinent~\cite{Nagle:2018nvi,ALICE:2017jyt,Adam:2016dau,Acharya:2018orn, Khachatryan:2010gv, Khachatryan:2016txc}. The interpretation of the results from the analysis of high-multiplicity \pp collisions is challenging due to the selection biases of the sample towards events in which partonic scatterings with large momentum transfer (hard scatterings) occurred. To mitigate this inherent bias, Martin \emph{et al.}~\cite{Martin:2016igp} suggested to use the charged-particle multiplicity in the transverse region ($N_{\mathrm{ch}}^{\mathrm{T}}$) as a classifier of the activity in the collisions, since the correlation between $N_{\mathrm{ch}}^{\mathrm{T}}$ and the hardest scattering in the collision is small. The ALICE collaboration has reported the first $N_{\mathrm{ch}}^{\mathrm{T}}$ spectra measured in pp collisions at centre-of-mass energy, $\sqrt{s}=13$\,\TeV~\cite{ALICE:2019mmy}. Event generators, such as \textsc{PYTHIA}~8~\cite{Sjostrand:2014zea} and EPOS-LHC~\cite{Pierog:2013ria}, do not provide a good description of the measured distribution of the ratio $N_{\mathrm{ch}}^{\mathrm{T}}$/$\langle N_{\mathrm{ch}}^{\mathrm{T}}\rangle$, where $\langle N_{\mathrm{ch}}^{\mathrm{T}}\rangle$ is the event-averaged charged-particle multiplicity in the transverse region, underestimating in particular the number of collisions with large $N_{\mathrm{ch}}^{\mathrm{T}}$ $\big(>3\times\langle N_{\mathrm{ch}}^{\mathrm{T}}\rangle\big)$. In the framework of MPI-based models, like those implemented in \textsc{PYTHIA}~8 and \textsc{HERWIG}~7~\cite{Bellm:2015jjp}, the probability for a hard scattering in the collision increases with decreasing impact parameter\footnote{In event generators like \textsc{PYTHIA}~8 the impact parameter profile is described by an overlap matter distribution of the two incoming hadrons.} between the colliding protons. Thus, requiring a high-\pt particle (e.g., $\ptt>8$\,\GeVc) in a given \pp collision biases the selection of collisions towards those with a smaller impact parameter~\cite{Strikman:2011cx}, which in turn biases the selection towards pp collisions with more MPI~\cite{ALICE:2019mmy}. This feature of the $N_{\mathrm{ch}}^{\mathrm{T}}$-based analysis is important for the isolation of potential MPI and colour reconnection effects, which according to \textsc{PYTHIA}~8, produce effects resembling collective behaviour~\cite{OrtizVelasquez:2013ofg}. By construction, MPI and colour reconnection effects are expected to be more relevant in the transverse region than in the away and toward regions~\cite{Ortiz:2018vgc}. It is worth mentioning that the MPI picture has been used to explain the \pt spectra in \pPb collisions and peripheral \PbPb collisions~\cite{ALICE:2014xsp,Loizides:2017sqq,ALICE:2018ekf}. Studies, as a function of $N_{\rm ch}^{\rm T}$, are therefore important to the understanding of the effects observed in high-multiplicity \pp collisions.  Last but not least, measurements of UE observables are also important to tune event generators~\cite{Skands:2014pea} that include hard partonic scatterings and MPI.

This letter reports the inclusive charged-particle transverse momentum spectra in \pp, \pPb and \PbPb collisions at centre-of-mass energy per nucleon pair $\sqrt{s_{\rm NN}}=5.02$\,\TeV containing a high-\pt leading particle within the kinematic intervals $8$~$\leq~\ptt <15$\,\GeVc and $|\eta|<0.8$. This guarantees the selection of collisions in which the average activity in the transverse region is roughly flat as a function of \ptt~\cite{ALICE:2019mmy}, and therefore, any additional selection on the charged particle multiplicity will only modulate the UE activity. The measurements are performed considering different event classes defined in terms of the multiplicity registered in the forward detectors. The \pt spectra of associated charged particles ($0.5 \leq p_{\rm T} < 6$\,\GeVc and $|\eta|<0.8$) are measured in the toward, away, and transverse regions as a function of the average charged particle multiplicity in the transverse region. 
To further investigate the possible modification of the particles produced in the hard scattering in \pp, \pPb, and \PbPb collisions,  the \pt distributions in the toward (${\rm{d}}N_{\rm{ch}}^{\rm{t}}/{\rm{d}}p_{\rm T}$) and away (${\rm{d}}N_{\rm{ch}}^{\rm{a}}/{\rm{d}}p_{\rm T}$) regions obtained after the subtraction of the \pt spectra in the transverse region (${\rm{d}}N_{\rm{ch}}^{\rm T}/{\rm{d}}p_{\rm T}$) are also reported. The subtracted yields (${\rm{d}}N_{\rm{ch}}^{\rm{st,sa}}/{\rm{d}}p_{\rm T}$) are further normalised to those measured in minimum-bias (MB) \pp collisions,
\begin{equation}
I_{X}^{\rm{t,a}} \equiv
\frac{({\rm{d}}N_{\rm{ch}}^{\rm{t,a}}/{\rm{d}}p_{\rm T} - {\rm{d}}N_{\rm{ch}}^{\rm T}/{\rm{d}}p_{\rm T})|_{X}}{ ({\rm{d}}N_{\rm{ch}}^{\rm{t,a}}/{\rm{d}}p_{\rm T} - {\rm{d}}N_{\rm{ch}}^{\rm T}/{\rm{d}}p_{\rm T})|_{\rm {pp,MB}}}=\frac{({\rm{d}}N_{\rm{ch}}^{\rm{st,sa}}/{\rm{d}}p_{\rm T})|_{X}}{ ({\rm{d}}N_{\rm{ch}}^{\rm{st,sa}}/{\rm{d}}p_{\rm T})|_{\rm {pp,MB}}},
\label{eq0}
\end{equation}
where $X$ indicates the collision system and the event multiplicity class.
In this way, the hard process \pt spectra in the toward and away regions are isolated, and thus allowing us to study possible modifications to the produced particles due to medium effects in high-multiplicity pp, \pPb, and \PbPb collisions. In heavy-ion collisions, this ratio is sensitive to the same effects which were studied using the $I_{\rm AA}$ quantity~\cite{Adams:2006yt,Aamodt:2011vg,ALICE:2016gso}, where jets produced in the early stage of the collision propagate through the hot and dense quark--gluon plasma. Their interaction with the coloured medium lead to parton-energy loss (jet quenching)~\cite{Qin:2015srf} which, for example, results in the suppression of the charged-particle yield at high \pt~\cite{ALICE:2018vuu}, and the suppression of the high-\pt yield in the away region~\cite{Adams:2006yt,Aamodt:2011vg}.  It is worth mentioning that jet quenching effects have not been observed so far in small systems~\cite{ALICE:2018vuu,ALICE:2017svf}.

\section{Experiment and data analysis} \label{sec:expSetup}

This analysis is based on the data recorded by the ALICE apparatus during the \pp and \PbPb runs at \snn = 5.02\,\TeV in 2015, and the \pPb run at \snn = 5.02\,\TeV in 2016. The present study uses the \VZERO detector, and the Silicon Pixel Detector (\SPD) for triggering and background rejection. The \VZERO consists of two arrays of scintillating tiles placed on each side of the interaction point covering the full azimuthal acceptance and the pseudorapidity intervals of 2.8~$<\eta<$~5.1 (\VZEROA) and -3.7~$<\eta<$~-1.7 (\VZEROC). The SPD is the innermost part of the Inner Tracking System (ITS) and it is the closest detector to the interaction point. It consists of two cylindrical silicon pixel layers at radial distances of 3.9 and 7.6 cm from the beam line and the pseudorapidity coverages of the two layers are $|\eta|<2$ and $|\eta|<1.4$, respectively.  The data were collected using a minimum-bias trigger, which required a signal in both \VZEROA and \VZEROC detectors. 
 The offline event selection was optimised to reject beam-induced background in all collision systems by utilising the timing signals in the two \VZERO detectors. In \PbPb collisions, the beam-induced background is further suppressed by correlating the timing signals of the neutron zero degree calorimeters, which are positioned on both sides of the interaction point at 112.5\,m distance along the beam axis~\cite{Abelev:2014ffa}. The signals from the zero degree calorimeters are also used to suppress the contamination from electromagnetic interactions. This is performed by requesting the coincidence of the signals coming from both side zero degree calorimeters by which the background due to single nucleus electromagnetic dissociation processes is excluded. A criterion based on the offline reconstruction of multiple primary vertices in the SPD is applied to reduce the pileup caused by multiple interactions in the same bunch crossing~\cite{ALICE:2008ngc}. The results presented in this letter are for minimum-bias triggered \pp collisions having at least one charged particle in the pseudorapidity interval $|\eta|<1$ (INEL$>$0).  The INEL$>$0 event class corresponds to about 75\% of the total inelastic cross section~\cite{Acharya:2019kyh}. For \pp and \PbPb collisions, the sample is subdivided into different multiplicity classes based on the total charge deposited in both \VZERO sub-detectors, which is termed as V0M amplitude~\cite{Adam:2015ptt}. For \pPb collisions, the sample is subdivided based on the total charge deposited in \VZEROA sub-detector (\VZEROA amplitude)~\cite{Adam:2014qja}, which is located in the Pb-going direction. The \VZEROA estimator has been implemented in previous measurements that used \pPb data (see e.g.~\cite{ALICE:2016dei}). This allows for comparisons with other observables for similar V0A multiplicity classes. To ensure that a hard scattering took place in the collision, events are required to have a trigger particle within $8 \leq p_{\rm T} ^{\rm trig} <15$ \GeVc. In this \ptt interval, the momentum resolution effects are negligible on the extracted yields, and therefore, no \ptt resolution correction is applied. The total number of analysed collisions before the trigger particle selection are about $10^8$, $10^8$, and $10^7$ for \pp, \pPb, and \PbPb collisions, respectively.

The transverse momentum of particles is determined from measurements in the central barrel with the ITS and the Time Projection Chamber (TPC). The ITS is a tracking detector which consists of six cylindrical layers of silicon detectors. The TPC is a cylindrical drift detector which covers a radial distance of 85-247\,cm from the beam axis and it has longitudinal dimension extending from about -250\,cm to +250\,cm around the nominal interaction point. Primary charged particles are measured in the pseudorapidity range of $|\eta|<0.8$ and with $p_{\rm T} > 0.5$\,\GeVc, where $\eta$ is measured in the laboratory frame for the three collision systems. The configuration for \pPb collisions with protons at 4\,\TeV energy colliding with Pb ions that have per-nucleon energies of ($Z/A$) $\times$ 4\,\TeV $\sim$ 1.58\,\TeV results in a shift in the rapidity of the nucleon–nucleon centre-of-mass system by 0.465 in the direction of the proton beam (negative z-direction). Here $Z$ and $A$ are the atomic and mass numbers of the Pb ion, respectively. Therefore, the detector coverage $|\eta|<0.8$ corresponds to roughly $-0.3<|\eta_{\rm{cms}}|<1.3$ for \pPb collisions. The particles with mean proper lifetime larger than 1\,cm/$c$, which are either produced directly in the interaction or from decays of particles with mean proper lifetime smaller than 1\,cm/$c$ are termed as primary particles~\cite{ALICE-PUBLIC-2017-005}. The track selection follows a procedure similar to the one described in Ref.~\cite{Acharya:2018qsh} and only few specific details are reported here.  Tracks ($N_{\rm{tracks}}$) are required to have two hits in the ITS, out of which at least one should be in either of the two innermost layers. The geometrical track length $L$ is calculated in the TPC readout plane, excluding the information from the pads at the sector boundaries ($\approx3$\,cm from the sector edges). The trajectory lengths built from radial segments, i.e. the crossed TPC pad rows, traversed in the TPC by a particle are required to be larger than 85\% of the geometrical track length. The pad rows are made of at least 3 neighbouring individual observations (clusters), and their height varies from 7.5\,mm to 15\,mm~\cite{Dellacasa:451098}. The trajectory lengths built from clusters (one cluster per pad row) is required to be larger than $0.7\times L$. The fraction of TPC clusters shared with another track is required to be lower than 0.4. The fit quality for the ITS and TPC track points must satisfy $\chi^{2}_{\mathrm{ITS}}/N_{\mathrm{hits}}<36$ and $\chi^{2}_{\mathrm{TPC}}/N_{\mathrm {clusters}}<4$, respectively, where $N_{\mathrm{hits}}$ and $N_{\mathrm{clusters}}$ are the numbers of hits in the ITS and the number of clusters in the TPC, respectively.
Only tracks with $\chi^{2}_{{\rm TPC}-{\rm ITS}}<36$ are included in the analysis, where $\chi^{2}_{{\rm TPC}-{\rm ITS}}$  is calculated comparing the track parameters from the combined ITS and TPC track reconstruction to that derived only from the TPC and constrained to the interaction point. The definition of $\chi^{2}_{{\rm TPC}-{\rm ITS}}$ can be found in Ref.~\cite{Abelev:2012hxa}. To reduce the contamination from secondary particles, tracks are accepted if their distance-of-closest-approach (DCA) to the reconstructed primary interaction vertex satisfies in the longitudinal ($d_{\rm z}$) and transverse ($d_{\rm xy}$) directions the conditions $d_{\rm z}<2$\,cm and $d_{\rm{xy}}<0.018$\,cm + $0.035$\,(cm$\times$\GeVc)$/p_{\rm T}$.

The measurement of the transverse momentum spectra of charged particles follows the standard procedure of the ALICE collaboration~\cite{Acharya:2019mzb,Acharya:2018qsh}. The raw yields are corrected for efficiency and contamination from secondary particles. The efficiency correction is calculated from Monte Carlo simulations with GEANT3~\cite{Brun:1082634} transport code, which made use of PYTHIA~8 (Monash)~\cite{Skands:2014pea}, EPOS-LHC~\cite{Pierog:2013ria} and HIJING~\cite{Deng:2010mv} event generators for \pp, \pPb and \PbPb collisions, respectively and incorporated a detailed description of the detector material, geometry and response.  Since the event generators do not reproduce the relative abundances of different particle species in the real data, the efficiency obtained from the simulations is re-weighted considering the particle composition from data as outlined in~\cite{Acharya:2018qsh}. A multi-component template fit based on the DCA distributions from the simulation is used for the estimation of secondary contamination~\cite{Acharya:2018qsh}.

The \pt spectra for the toward and away regions include contributions from the jet fragmentation, ISR, and FSR, as well as, the contribution from the underlying event. In order to increase the sensitivity to the hardest process of the event, the particle yields measured in the transverse region are subtracted from the corresponding yields in both the toward and away regions: ${\rm{d}}N_{\rm{ch}}^{\rm{t,a}}/{\rm{d}}p_{\rm T} - {\rm{d}}N_{\rm{ch}}^{\rm T}/{\rm{d}}p_{\rm T}$. This approach assumes that the background (UE, ISR, and FSR) in the toward and away regions is similar to the activity in the transverse region. However, one has to keep in mind that in \PbPb collisions two-particle correlations are affected by anisotropic transverse flow. In particular, the main contribution is due to the elliptic flow, $v_2$, which is the second order coefficient in the Fourier expansion of the azimuthal distribution of the particle momenta~\cite{ALICE:2011svq}. This elliptic azimuthal anisotropy modulates the background according to:
\begin{equation}
B(\Delta\varphi)=B_{0}\big(  1 + 2V_{2}\cos{(2\Delta\varphi)} \big),
\label{eqv2}
\end{equation}
where $V_{2}$ is approximately given by the product of anisotropic flow coefficients for trigger and associated particles at their respective momenta i.e. $V_{2}\approx v_{2}^{\rm trig}v_{2}^{\rm assoc}$. The existing $v_{2}$ measurements over a broad transverse momentum range~\cite{ALICE:2012vgf} suggest that the effect of the $v_2$ modulation of background should be more relevant in semi-central \PbPb collisions. The effect is expected to be important at low and intermediate transverse momenta and decreases for high transverse momentum particles~\cite{Aamodt:2011vg}. In the high-\pt region of interest for the jet quenching studies, namely \pt$> 4$\,\GeVc,  the effect of the $v_2$ modulation is estimated to be small (about 5\%) for \PbPb collisions. Given that the $v_2$ effect is larger in \PbPb collisions than in \pp and \pPb collisions, no correction for the $v_2$ modulation is applied for \pp and \pPb collisions since its effect is smaller than the other sources of systematic uncertainty. 

The results are shown as a function of the average number of charged particles in the transverse region $\langle N_{\rm ch}^{\rm T} \rangle$. The values of $\langle N_{\rm ch}^{\rm T} \rangle$ are extracted in each multiplicity class from the $N_{\rm{tracks}}$~distributions in the transverse region that are corrected for detector effects using a Bayesian unfolding~\cite{DAgostini:1994fjx}. The Bayesian unfolding requires the multiplicity response matrix, which is built from the correlation between the measured multiplicity and the multiplicity at generator level (without detector effects) in the transverse region. This has been obtained from MC simulations which include the propagation of particles through the detector using GEANT~3. As a crosscheck, the $\langle N_{\rm ch}^{\rm T} \rangle$ values are also calculated by integrating the transverse momentum distributions in the interval $0.5 \leq p_{\rm T} < 8$\,\GeVc. The difference between the results from the two strategies is assigned as the systematic uncertainty on $\langle N_{\rm ch}^{\rm T} \rangle$, where the effects related to the discrepancy between data and MC in the particle composition and secondary contamination are considered. This uncertainty amounts up to 3.5\%, 4\% and 6.5\% for \pp, \pPb and \PbPb collisions, respectively.

The systematic uncertainties related to the track selection criteria were studied by repeating the analysis varying one-by-one the track selection criteria~\cite{Acharya:2018qsh,Acharya:2019mzb}. In particular, the upper limits of the track fit quality parameters in the ITS ($\chi^2_{\rm {ITS}}/N_{\rm {hits}}$) and in the TPC ($\chi^2_{\rm {TPC}}/N_{\rm{clusters}}$) were varied in the ranges of 25--49 and 3--5, respectively. The maximum fraction of shared TPC clusters was varied between 0.2 to 1 and the maximum $d_{\rm z}$ was varied between 1 and 5\,cm~\cite{Acharya:2018qsh}. We have also quantified the impact of not including the ITS hit requirement in the track selection. The systematic uncertainty on the primary particle composition was estimated using a procedure similar to the one described in~\cite{Acharya:2018qsh}. To quantify the uncertainty due to the imperfect simulation of the detector response, the track matching between the TPC and the ITS information in the data and in the simulation were compared. To achieve this, the fraction of secondary particles was rescaled according to fits to the measured DCA distributions. After this rescaling, the agreement between data and model was found to be within 3\% for all collision systems. This value was assigned as an additional systematic uncertainty~\cite{Acharya:2018qsh}. 
\begin{table} [!hpt]
                \centering 
                \caption{Contributions to the relative (\%) systematic uncertainty on the \pt spectra of primary charged particles in \pp, \pPb, and \PbPb collisions at \snn $=$ 5.02\,\TeV. Just for illustration, the range in the table corresponds to the lowest and highest relative systematic uncertainty in the considered \pt range. The individual contributions are summed in quadrature to obtain the total uncertainty.
                 \label{tab:sys}}
                \begin{tabular}{|l |l |l |l |}
                \hline
                \textbf{Source of uncertainty}  & \pp & \pPb & \PbPb\\ 
                \hline
                Track selection                 &2.1--8.2  &2.4--5.8 &3.0--9.9 \\
                Particle composition            &0.3--1.8 &0.5--1.9     &0.3--2.4 \\
                Secondary particles & 0.0--0.4 &0.0--2.4 &0.0--1.9 \\
                Matching efficiency    &2.0--4.2&0.7--3.7&0.6--3.7 \\
                \hline
                Total               &3.2--8.8 &3.6--6.3  &3.5--10.0 \\
                \hline
                Total ($N_{\rm ch}$-dependent)               &2.0--4.5  &1.7--4.0 &1.1--3.7 \\
                \hline
                \end{tabular}
               
\end{table}
The systematic uncertainty on the secondary particle contamination considers the imperfection of the method (multi-component template fit) used to extract the correction. The fit ranges were varied and the fit was repeated using templates with two (primaries, secondaries) or three (primaries, secondaries  from material, secondaries from weak decays) components. The maximum spread among these variations was assigned as the systematic uncertainty on the secondary contamination. This contribution dominates at low $p_{\rm T}$. The density of materials used in simulations of the experimental setup was varied by $\pm$ 4.5\%~\cite{Abelev:2014ffa}, resulting in a negligible systematic uncertainty in the considered $p_{\rm T}$ range of 0.5 to 6.0 \GeVc. For the estimation of total systematic uncertainty, all the above listed contributions were summed in quadrature. The systematic uncertainties are independent of the difference between the azimuthal angle of the associated particle and that of the trigger particle. The estimated systematic uncertainties on the \pt spectra significantly depend on $p_{\rm T}$, while the dependence on the multiplicity classes is mild. The ranges of systematic uncertainties in the three considered collision systems are reported in Table~\ref{tab:sys} for the various sources described above.

\section{Results and discussion}\label{sec:results}

\begin{figure}[ht!]
\centering
\includegraphics[scale=0.8]{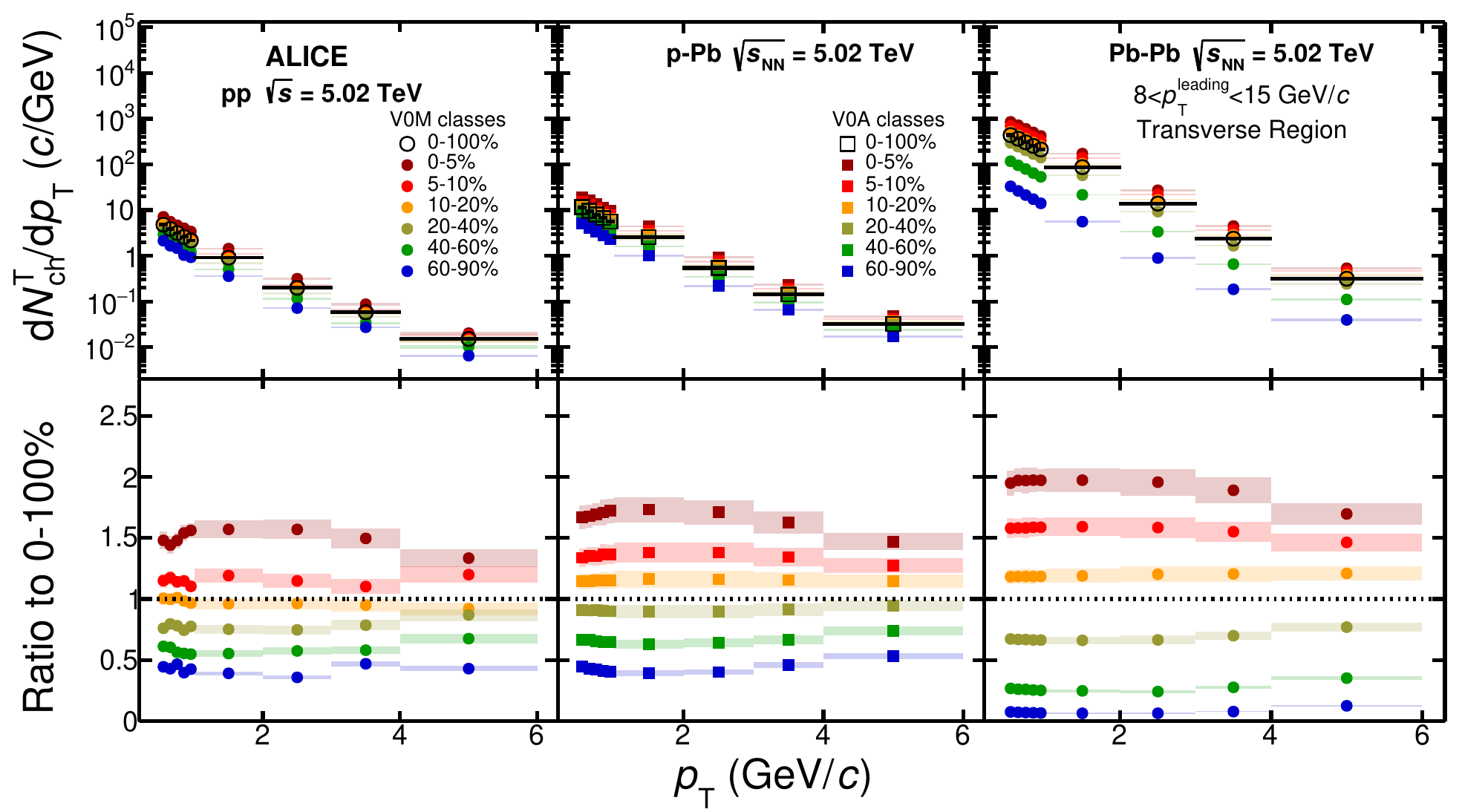}
\caption{(colour online) Top panels: transverse momentum spectra of charged particles in the transverse region for different multiplicity classes in \pp (left), \pPb (middle) and \PbPb (right) collisions at $\sqrt{s_{\rm NN}}=5.02$\,\TeV. The \pt spectra are measured at mid pseudorapidity ($|\eta|<0.8$). Lower panels: Ratio of \pt spectra in different multiplicity classes to the \pt spectrum in the 0$-$100\% multiplicity class for the corresponding collision systems. The statistical and systematic uncertainties are shown by bars and boxes, respectively.
}
\label{fig1}
\end{figure}

The \pt spectra measured in the transverse region for \pp, \pPb, and \PbPb collisions are shown in Fig.~\ref{fig1} (top panel). Results are presented for different multiplicity classes. The ratios between the spectra in the individual multiplicity classes and the MB (0$-$100\%) one are shown in the bottom panel. In the \pt range $0.5-6$\,GeV/$c$, the ratios for the highest multiplicity class (0$-$5\%) are larger than unity and show an increasing trend with increasing \pt at low \pt ($<2-3$\,\GeVc) followed at higher \pt by a slow decrease. Instead, for the lowest multiplicity classes (40$-$60\% and 60$-$90\%) the ratios are lower than unity and follow an opposite trend with $p_{\rm T}$, decreasing at low \pt and increasing for \pt$>3$\,\GeVc. The behaviour of the ratios as a function of the event activity is reminiscent of analogous ratios as a function of the number of MPI in \pp collisions simulated with \textsc{PYTHIA}~8, including colour reconnection~\cite{Ortiz:2020rwg}. In particular, at $p_{\rm T} \approx 2-3$ GeV/$c$ the \pt spectrum of pp collisions with large MPI activity exhibits an enhancement with respect to the \pt spectrum of MB pp collisions. The effect was not observed before in data because, in contrast to the present analysis, the jet contribution was included in the \pt spectra~\cite{Acharya:2019mzb}.

\begin{figure}[ht!]
\centering
\includegraphics[width=37pc]{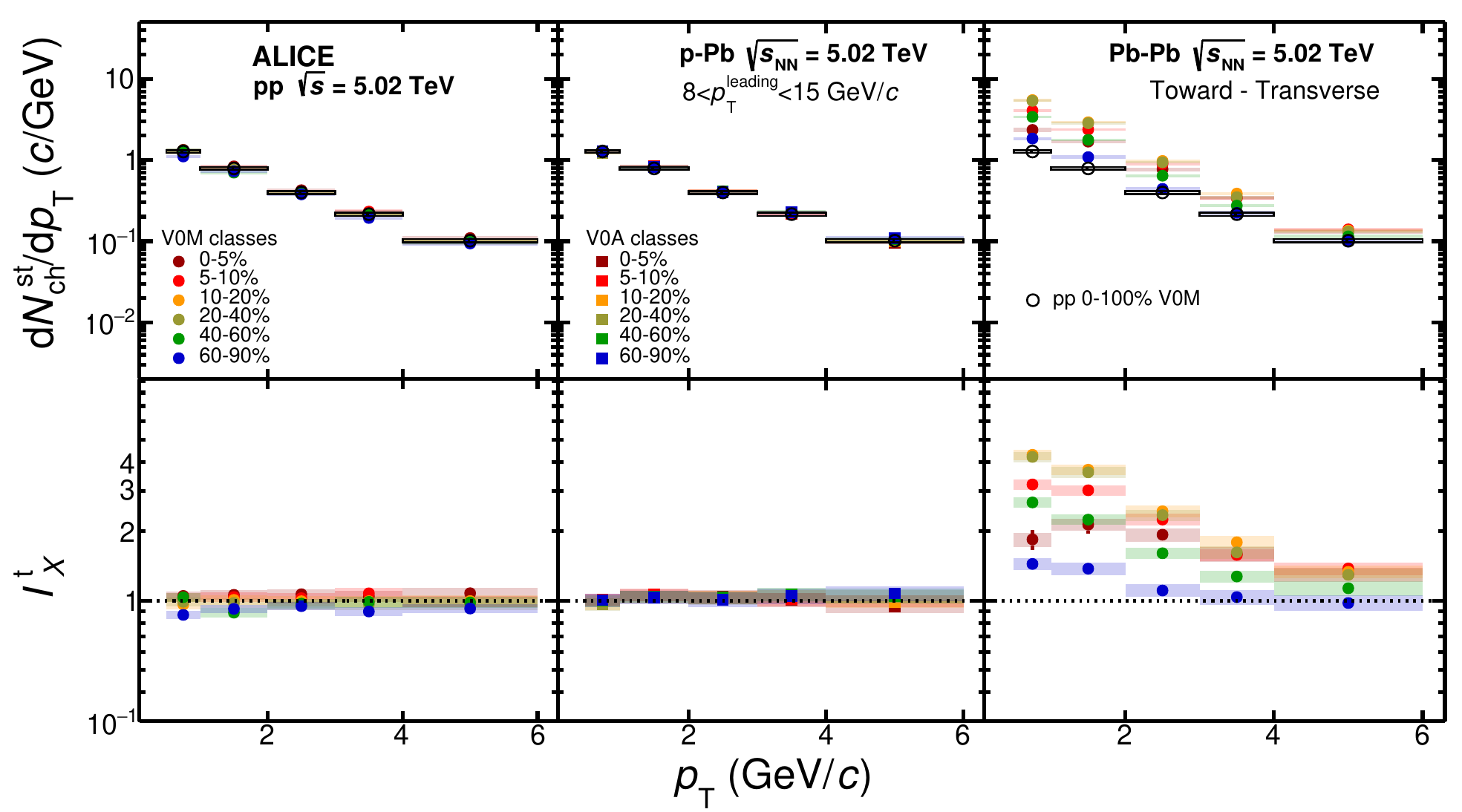}
\includegraphics[width=37pc]{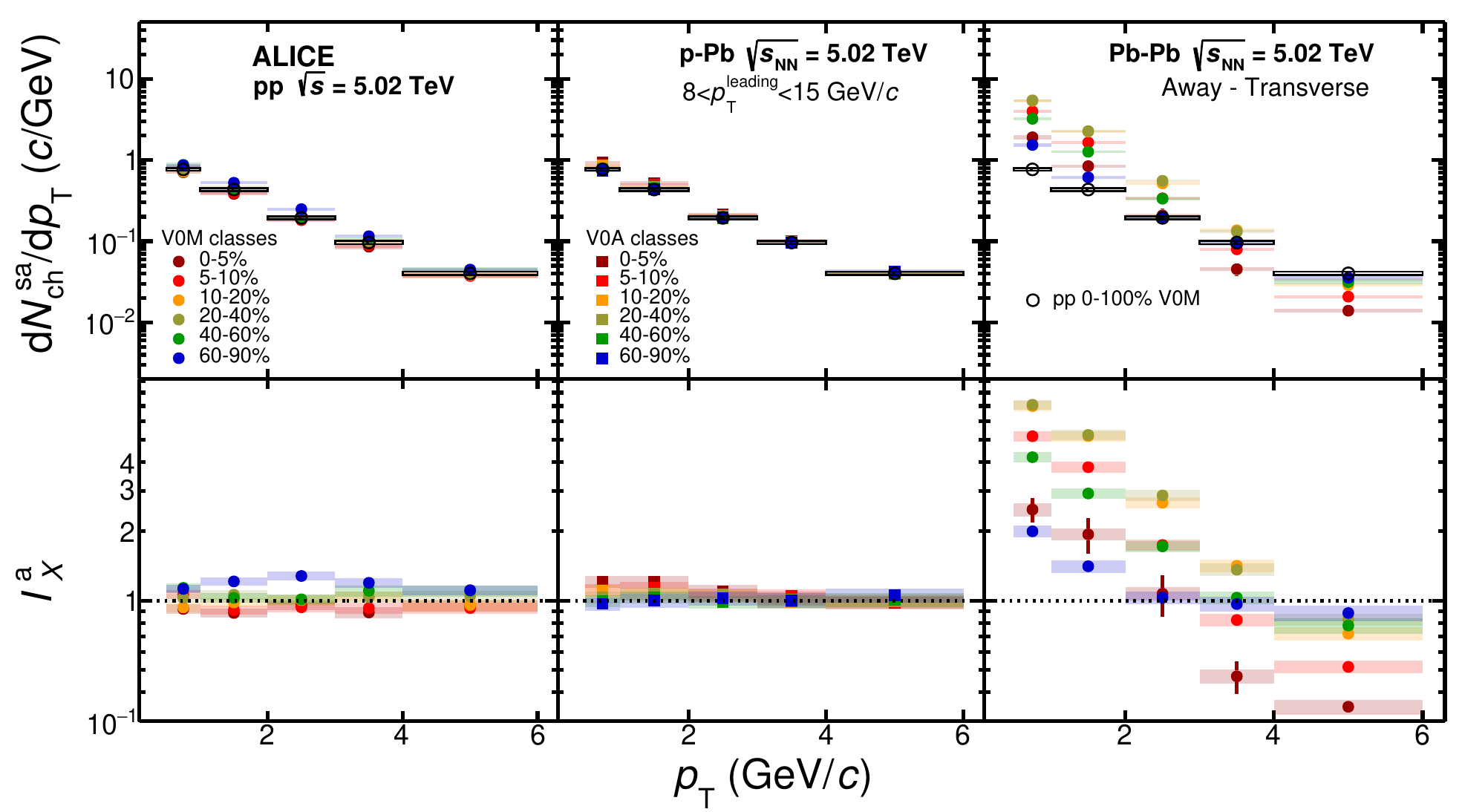}
\caption{(colour online) Transverse momentum spectra of charged particles in Toward-Transverse, ${\rm{d}}N_{\rm{ch}}^{\rm{st}}/{\rm{d}}p_{\rm T}$ (top plot) and Away-Transverse, ${\rm{d}}N_{\rm{ch}}^{\rm{sa}}/{\rm{d}}p_{\rm T}$ (bottom plot) regions for different multiplicity classes in \pp (left), \pPb (middle) and \PbPb (right) collisions at $\sqrt{s_{\rm NN}}=5.02$\,\TeV. The \pt spectra are measured at mid pseudorapidity ($|\eta|<0.8$). The lower panels of both plots show the ratio to minimum bias \pp collisions. The statistical and systematic uncertainties are shown by bars and boxes, respectively.
}
\label{fig2}
\end{figure}

The top (bottom) panel of Fig.~\ref{fig2} shows the charged particle yields for the toward (away) region after the subtraction of the yields measured in the transverse region in \pp, \pPb and \PbPb collisions. Results are compared with the \pt spectra measured for MB pp collisions (0$-$100\% V0M \pp event class) quantified with the ratio $I_{X}^{\rm{t,a}}$, as defined in Eq.~\ref{eq0}. At low transverse momenta, $p_{\rm T}<4$\,\GeVc, $I_{X}^{\rm{t,a}}$ is close to unity in \pp and \pPb collisions. In contrast, $I_{X}^{\rm{t,a}}$ in \PbPb collisions exhibits a strong multiplicity dependence over the whole measured \pt interval. The $I_{X}^{\rm{t,a}}$ magnitude is larger for semi-peripheral \PbPb collisions, the maximum is observed for 20$-$40\% \PbPb collisions, and is smaller for the most central and most peripheral classes. Given that the $v_2$ contribution is not subtracted from the jet-like yields reported in Fig.~\ref{fig2}, the centrality dependence of $I_{X}^{\rm{t,a}}$ follows the behaviour of $v_2$ as a function of collision centrality and particle \pt in \PbPb collisions at LHC energies~\cite{Adam:2016izf}.

\begin{figure}[ht!]
\centering
\includegraphics[width=37pc]{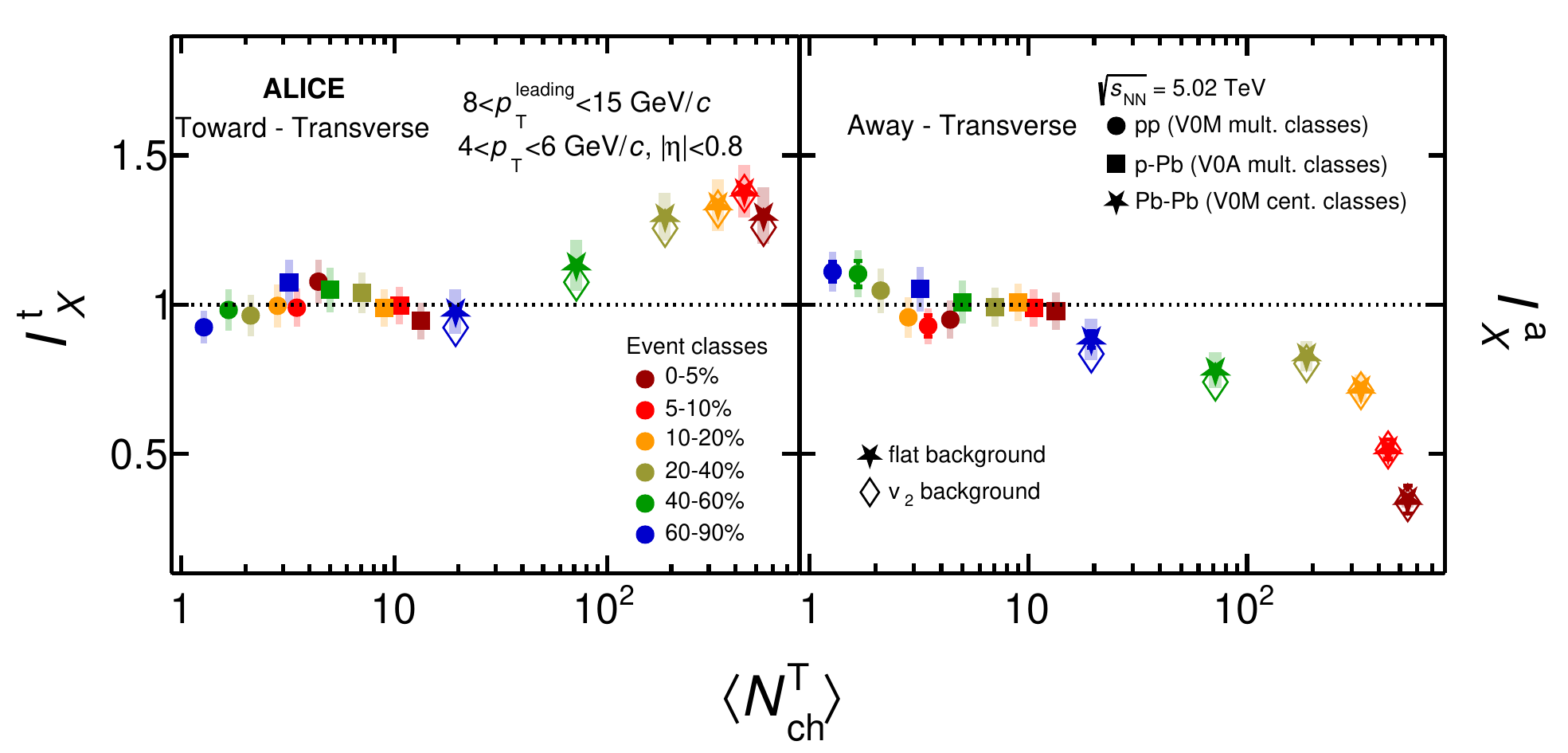}
\caption{(colour online) The $I_{X}^{\rm t}$ (left) and $I_{X}^{\rm a}$ (right) as a function of $\langle N_{\rm ch}^{\rm T} \rangle$ in 4 $< p_{\rm T} < $ 6 \GeVc for different multiplicity classes in \pp, \pPb and \PbPb collisions at \snn $=$ 5.02\,\TeV. \PbPb results are shown assuming a flat background (filled markers), and assuming a $v_{2}$-modulated background (empty markers). The statistical and systematic uncertainties are shown by bars and boxes, respectively.}
\label{fig3}
\end{figure}

Figure~\ref{fig3} shows the measured values of $I_{\rm{X}}^{\rm{t,a}}$ in the transverse momentum interval $4<p_{\rm T}<6$\,\GeVc as a function of the average multiplicity in the transverse region for all the multiplicity classes considered in \pp, \pPb and \PbPb collisions.
The figure shows that, within uncertainties, the $I_{\rm{X}}^{\rm{t,a}}$ values are close to unity for all the multiplicity classes measured in \pp and \pPb collisions. This indicates that effects induced by possible energy loss in these systems are not observed within uncertainties. This result is consistent with previous studies of nuclear modification factor~\cite{ALICE:2018vuu} and hadron-jet recoil measurements~\cite{ALICE:2017svf}.
By contrast, for \PbPb collisions the $I_{\rm{X}}^{\rm{t,a}}$ values are compatible to unity for peripheral collisions, and show a gradual enhancement (reduction) with the increase in multiplicity for the toward (away) region. The behaviour is the same for the $I_{\rm{X}}^{\rm{t,a}}$ values measured either assuming a flat background or a $v_2$-modulated background. The $v_2$-modulated background was estimated following the approach depicted in Eq.~\ref{eqv2} and using the $v_2$ data reported in~\cite{ALICE:2012vgf}. This behaviour is qualitatively similar to the di-hadron correlation results reported by the STAR and ALICE collaborations~\cite{Adams:2006yt,Aamodt:2011vg}. In \PbPb collisions, $I_{\rm{X}}^{\rm{t}}$ provides information about the fragmenting jet leaving the medium, while on the away side, $I_{\rm{X}}^{\rm{a}}$ reflects the survival probability of the recoiling parton during passage through the medium. Thus a suppression of $I_{\rm{X}}^{\rm{a}}$ would indicate that fewer partons survive the passage through the medium and is expected from the strong in-medium energy loss. 
On the other hand, the enhancement observed in the toward region is also subject to medium effects. The ratio is sensitive to a) a possible change of the fragmentation functions, b) a possible modification of the quark to gluon jet ratio in the final state due to different coupling with medium, and c) a possible bias on the parton spectrum due to trigger particle selection. Moreover, given that $I_{X}^{\rm{t,a}}$ is sensitive to the same effects as $I_{\rm AA}$, the interpretation of the results is similar to that reported in~\cite{Aamodt:2011vg}.
It is likely that all three effects play a role~\cite{Aamodt:2011vg}. A detailed quantification of the contribution of each effect is beyond the scope of the present paper.

\begin{figure}[ht!]
\centering
\includegraphics[width=37pc]{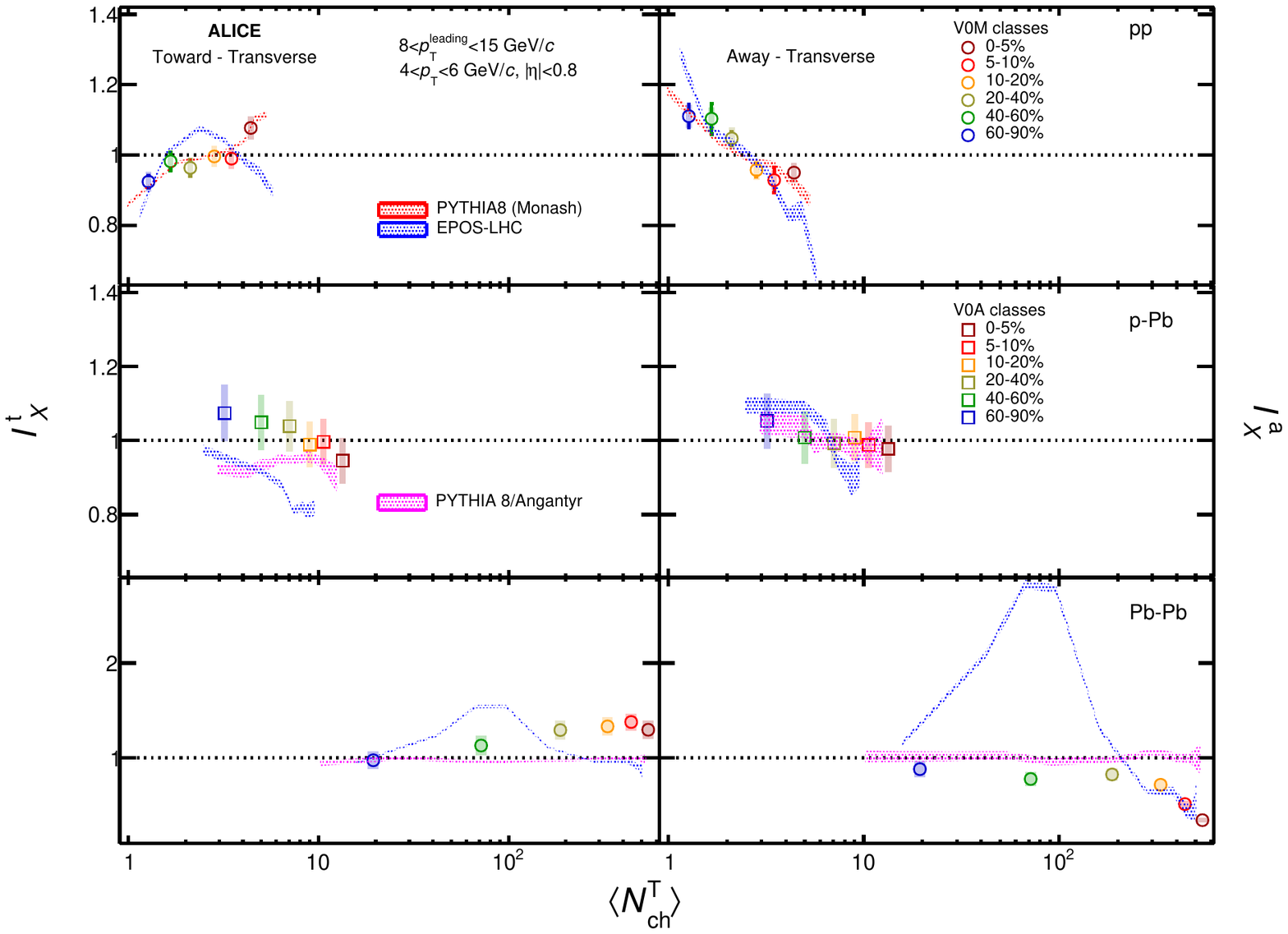}
\caption{(colour online) Comparison of the measured the $I_{X}^{\rm t}$ (left) and $I_{X}^{\rm a}$ (right) in 4 $< p_{\rm T} < $ 6 \GeVc with model predictions. The results are shown as a function of $\langle N_{\rm ch}^{\rm T} \rangle$ for different multiplicity classes in \pp (top panel), \pPb (middle panel) and \PbPb (bottom panel) collisions at \snn $=$ 5.02\,\TeV. The red and magenta lines show the \textsc{PYTHIA}~8 (Monash)~\cite{Skands:2014pea} and \textsc{PYTHIA}~8/Angantyr~\cite{Skands:2014pea} predictions, respectively. The blue lines show the EPOS-LHC~\cite{Pierog:2013ria} results. The statistical and systematic uncertainties are shown by bars and boxes, respectively.}
\label{fig4}
\end{figure}

In order to get further insight into the effect, the measured $I_{X}^{\rm{t,a}}$ values are compared in Fig.~\ref{fig4} with model predictions. Following the similar treatment of the experimental data, for the models, the total sample is subdivided into different V0M classes and the $\langle N_{\rm ch}^{\rm T} \rangle$ is calculated for each class. 
For high-multiplicity \pp collisions, although $I_{X}^{\rm{t,a}}$ is close to unity, a small trend with multiplicity is visible, which is not seen at similar multiplicities (20$-$90\% V0A) in \pPb data. To understand the source of these slight deviations from unity, the data are compared with the predictions from the \textsc{PYTHIA}~8 (Monash tune~\cite{Skands:2014pea}) and EPOS-LHC~\cite{Pierog:2013ria} event generators.
In PYTHIA, the hadronization of quarks is simulated using the Lund string fragmentation model~\cite{Andersson:1983ia}. Various PYTHIA tunes have been developed through extensive comparison of Monte Carlo distributions with the minimum-bias data from different experiments. The Monash tune of \textsc{PYTHIA}~8 is tuned to LHC data and uses an updated set of hadronization parameters compared to the previous tunes~\cite{Skands:2014pea}. EPOS-LHC is built on the Parton-Based Gribov Regge Theory. Utilising the colour exchange mechanism of string excitation, the model is tuned to LHC data~\cite{Pierog:2013ria}. In this model, a part of the collision system which has high parton densities becomes a ``core" region that evolves hydrodynamically as a quark--gluon plasma and it is surrounded by a more
dilute ``corona" for which fragmentation occurs  in the vacuum. 
The upper panel of Fig.~\ref{fig4} shows $I_{X}^{\rm{t,a}}$ for different multiplicity classes. The observed deviations from unity are reproduced by \textsc{PYTHIA}~8 for both the toward and away regions. Given that \textsc{PYTHIA}~8 does not incorporate any jet quenching mechanism, the origin of the effect in high $\langle N_{\rm ch}^{\rm T} \rangle$ collisions is related to a remaining bias towards harder fragmentation and more activity from initial and final state radiation~\cite{Bencedi:2021tst}. These effects enhance the high-\pt yield in the toward region, and produce a broadening in the away region~\cite{Ortiz:2020dph}. The EPOS-LHC results in the away region are similar to both data and \textsc{PYTHIA}~8. However, for $I_{X}^{\rm{t}}$ EPOS-LHC exhibits a trend with a maximum at intermediate multiplicity and a reduction toward low and high multiplicities, which is not consistent with the measurements. 

The middle and bottom panels of Fig.~\ref{fig4} show $I_{X}^{\rm{t,a}}$ measured for \pPb and \PbPb collisions, respectively. The data are compared to \textsc{PYTHIA}~8/Angantyr~\cite{Bierlich:2018xfw} and EPOS-LHC predictions. The Angantyr model in \textsc{PYTHIA}~8 extrapolates the dynamics from \pp collisions to \pPb and \PbPb collisions, generalising the formalism adopted for \pp collisions by including a description of the nucleon positions within the colliding nuclei and utilising the Glauber model to calculate the number of interacting nucleons and binary nucleon--nucleon collisions.
\textsc{PYTHIA}~8/Angantyr, which does not include jet quenching effects, predicts $I_{X}^{\rm{t,a}}$ values consistent with unity for all the multiplicity classes in \PbPb collisions. Whereas for \pPb collisions $I_{X}^{\rm{a}}$ is consistent with unity, and $I_{X}^{\rm{t}}$ is slightly below unity. In EPOS-LHC, a certain \pt cutoff is defined in such a way that, above this cutoff, a particle loses part of its momentum in the core but survives as an independent particle produced by a flux tube. Soft particles, which are below the \pt cutoff, get completely absorbed and form the core. This sort of energy loss mechanism implemented in EPOS-LHC depends on the system size~\cite{Pierog:2013ria,Baier:1996kr,Peigne:2008ns}. Figure~\ref{fig4} (middle) shows that for \pPb collisions, EPOS-LHC does not describe either the magnitude or the trend of the multiplicity dependence of the measured ratio in the toward region, $I_{X}^{\rm{t}}$. However, the model is in reasonable agreement with data in the away region. For \PbPb collisions, EPOS-LHC predicts a significant enhancement of $I_{X}^{\rm{t,a}}$ for low $\langle N_{\rm ch}^{\rm T} \rangle$ ranges and deviates significantly from the experimental results. 

In summary, while the data from \PbPb collisions are in qualitative agreement with expectations from parton energy loss due to the presence of a hot and dense medium, \pp and \pPb data do not show any hint of medium effects in the multiplicity range which is reported.

\section{Summary}\label{sec:conc}
The transverse momentum spectra ($0.5\leq p_{\rm T}<6$\,\GeVc) of primary charged particles in three azimuthal regions (toward, away and transverse) defined with respect to the direction of the particle with the highest transverse momentum in the event ($8\leq p_{\rm T}^{\rm trig}<15$\,\GeVc) are reported. The spectra are studied in intervals of the multiplicity measured at forward pseudorapidities for \pp, \pPb, and \PbPb collisions at $\sqrt{s_{\rm NN}}=5.02$\,\TeV.
The \pt spectra in the transverse region are subtracted from those of  the away and toward regions. This is based on the assumption that the transverse side provides a good estimation of the underlying event contribution in both the toward and away regions. However, for the interpretation of the results one has to keep in mind that $v_2$ modulates the background and this effect is important for semi-central \PbPb collisions and for $p_{\rm T}>4$\,\GeVc the effect is less than 5\% in central and peripheral \PbPb collisions. Ratios to MB \pp ($I_{X}^{\rm{t,a}}$), i.e., the multiplicity dependent yields normalised to the yield measured in MB pp collisions, are reported. At low transverse momentum ($p_{\rm T}< 2$\,\GeVc), within 20\%, the $I_{X}^{\rm{t,a}}$ values are multiplicity independent for both the toward and away regions in \pp and \pPb collisions. In contrast, in \PbPb collisions for both toward and away regions the $I_{X}^{\rm{t,a}}$  values exhibit a centrality dependence which is expected given the residual presence of elliptic flow. In the highest transverse momentum interval ($4<p_{\rm T}<6$\,\GeVc), the $I_{X}^{\rm{t,a}}$ values in \pp collisions are closer to unity but they exhibit a small reduction (increase) towards high \VZERO activity in \pp collisions. This trend is well reproduced by \textsc{PYTHIA}~8. In the model, it is due to a selection bias towards \pp collisions with harder fragmentation and larger activity from initial and final state radiation. For \pPb collisions, within uncertainties, the $I_{X}^{\rm{t,a}}$ values are consistent with unity and do not show a multiplicity dependence. \textsc{PYTHIA}~8/Angantyr fairly describes $I_{X}^{\rm{a}}$, but it underestimates by about 10\% the $I_{X}^{\rm{t}}$ values in the low multiplicity classes (40$-$90\% \VZEROA event class). For \PbPb collisions, the $I_{X}^{\rm{t,a}}$ values are close to unity for peripheral collisions, and show a gradual increase (reduction) in the toward (away) region with increasing multiplicity. A similar observable, $I_{\rm{AA}}$, based on the per-trigger yield of associated particles in di-hadron correlation has been studied for central and peripheral \PbPb collisions at $\sqrt{s_{\rm NN}}=2.76$\,TeV. The behaviour of $I_{X}^{\rm{t,a}}$ exhibits the same features as $I_{\rm{AA}}$: in central collisions, on the away-side, a suppression is observed as expected from strong in-medium energy loss. In the toward region, an enhancement is observed.
\textsc{PYTHIA}~8/Angantyr predicts $I_{X}^{\rm{t,a}} \approx 1$ for all multiplicity intervals, and it does not reproduce the observed away-side suppression or toward-side enhancement. Generally, EPOS-LHC does not describe the measured $I_{X}^{\rm{t,a}}$ ratios.

In summary, within the multiplicity reach reported in this paper, no jet quenching effects are observed in \pp and \pPb collisions within uncertainties. Further studies are required to extend the present analysis to higher multiplicities, which are currently limited by the event selection based on the forward \VZERO detector. The analysis of future \pp and \pPb collisions with much larger integrated luminosity may remove this limitation.

\newenvironment{acknowledgement}{\relax}{\relax}
\begin{acknowledgement}
\section*{Acknowledgements}

The ALICE Collaboration would like to thank all its engineers and technicians for their invaluable contributions to the construction of the experiment and the CERN accelerator teams for the outstanding performance of the LHC complex.
The ALICE Collaboration gratefully acknowledges the resources and support provided by all Grid centres and the Worldwide LHC Computing Grid (WLCG) collaboration.
The ALICE Collaboration acknowledges the following funding agencies for their support in building and running the ALICE detector:
A. I. Alikhanyan National Science Laboratory (Yerevan Physics Institute) Foundation (ANSL), State Committee of Science and World Federation of Scientists (WFS), Armenia;
Austrian Academy of Sciences, Austrian Science Fund (FWF): [M 2467-N36] and Nationalstiftung f\"{u}r Forschung, Technologie und Entwicklung, Austria;
Ministry of Communications and High Technologies, National Nuclear Research Center, Azerbaijan;
Conselho Nacional de Desenvolvimento Cient\'{\i}fico e Tecnol\'{o}gico (CNPq), Financiadora de Estudos e Projetos (Finep), Funda\c{c}\~{a}o de Amparo \`{a} Pesquisa do Estado de S\~{a}o Paulo (FAPESP) and Universidade Federal do Rio Grande do Sul (UFRGS), Brazil;
Bulgarian Ministry of Education and Science, within the National Roadmap for Research Infrastructures 2020-2027 (object CERN), Bulgaria;
Ministry of Education of China (MOEC) , Ministry of Science \& Technology of China (MSTC) and National Natural Science Foundation of China (NSFC), China;
Ministry of Science and Education and Croatian Science Foundation, Croatia;
Centro de Aplicaciones Tecnol\'{o}gicas y Desarrollo Nuclear (CEADEN), Cubaenerg\'{\i}a, Cuba;
Ministry of Education, Youth and Sports of the Czech Republic, Czech Republic;
The Danish Council for Independent Research | Natural Sciences, the VILLUM FONDEN and Danish National Research Foundation (DNRF), Denmark;
Helsinki Institute of Physics (HIP), Finland;
Commissariat \`{a} l'Energie Atomique (CEA) and Institut National de Physique Nucl\'{e}aire et de Physique des Particules (IN2P3) and Centre National de la Recherche Scientifique (CNRS), France;
Bundesministerium f\"{u}r Bildung und Forschung (BMBF) and GSI Helmholtzzentrum f\"{u}r Schwerionenforschung GmbH, Germany;
General Secretariat for Research and Technology, Ministry of Education, Research and Religions, Greece;
National Research, Development and Innovation Office, Hungary;
Department of Atomic Energy Government of India (DAE), Department of Science and Technology, Government of India (DST), University Grants Commission, Government of India (UGC) and Council of Scientific and Industrial Research (CSIR), India;
National Research and Innovation Agency - BRIN, Indonesia;
Istituto Nazionale di Fisica Nucleare (INFN), Italy;
Japanese Ministry of Education, Culture, Sports, Science and Technology (MEXT) and Japan Society for the Promotion of Science (JSPS) KAKENHI, Japan;
Consejo Nacional de Ciencia (CONACYT) y Tecnolog\'{i}a, through Fondo de Cooperaci\'{o}n Internacional en Ciencia y Tecnolog\'{i}a (FONCICYT) and Direcci\'{o}n General de Asuntos del Personal Academico (DGAPA), Mexico;
Nederlandse Organisatie voor Wetenschappelijk Onderzoek (NWO), Netherlands;
The Research Council of Norway, Norway;
Commission on Science and Technology for Sustainable Development in the South (COMSATS), Pakistan;
Pontificia Universidad Cat\'{o}lica del Per\'{u}, Peru;
Ministry of Education and Science, National Science Centre and WUT ID-UB, Poland;
Korea Institute of Science and Technology Information and National Research Foundation of Korea (NRF), Republic of Korea;
Ministry of Education and Scientific Research, Institute of Atomic Physics, Ministry of Research and Innovation and Institute of Atomic Physics and University Politehnica of Bucharest, Romania;
Ministry of Education, Science, Research and Sport of the Slovak Republic, Slovakia;
National Research Foundation of South Africa, South Africa;
Swedish Research Council (VR) and Knut \& Alice Wallenberg Foundation (KAW), Sweden;
European Organization for Nuclear Research, Switzerland;
Suranaree University of Technology (SUT), National Science and Technology Development Agency (NSTDA), Thailand Science Research and Innovation (TSRI) and National Science, Research and Innovation Fund (NSRF), Thailand;
Turkish Energy, Nuclear and Mineral Research Agency (TENMAK), Turkey;
National Academy of  Sciences of Ukraine, Ukraine;
Science and Technology Facilities Council (STFC), United Kingdom;
National Science Foundation of the United States of America (NSF) and United States Department of Energy, Office of Nuclear Physics (DOE NP), United States of America.
In addition, individual groups or members have received support from:
Marie Sk\l{}odowska Curie, Strong 2020 - Horizon 2020, European Research Council (grant nos. 824093, 896850, 950692), European Union;
Academy of Finland (Center of Excellence in Quark Matter) (grant nos. 346327, 346328), Finland;
Programa de Apoyos para la Superaci\'{o}n del Personal Acad\'{e}mico, UNAM, Mexico.
   
      
\end{acknowledgement}
\bibliographystyle{utphys}
\bibliography{biblio2}

\newpage
\appendix
    
\section{The ALICE Collaboration}
\label{app:collab}
\begin{flushleft} 
\small

S.~Acharya\,\orcidlink{0000-0002-9213-5329}\,$^{\rm 124,131}$, 
D.~Adamov\'{a}\,\orcidlink{0000-0002-0504-7428}\,$^{\rm 86}$, 
A.~Adler$^{\rm 69}$, 
G.~Aglieri Rinella\,\orcidlink{0000-0002-9611-3696}\,$^{\rm 32}$, 
M.~Agnello\,\orcidlink{0000-0002-0760-5075}\,$^{\rm 29}$, 
N.~Agrawal\,\orcidlink{0000-0003-0348-9836}\,$^{\rm 50}$, 
Z.~Ahammed\,\orcidlink{0000-0001-5241-7412}\,$^{\rm 131}$, 
S.~Ahmad\,\orcidlink{0000-0003-0497-5705}\,$^{\rm 15}$, 
S.U.~Ahn\,\orcidlink{0000-0001-8847-489X}\,$^{\rm 70}$, 
I.~Ahuja\,\orcidlink{0000-0002-4417-1392}\,$^{\rm 37}$
% ...

\section*{Affiliation Notes}

$^{\rm I}$ Deceased\\
$^{\rm II}$ Also at: Max-Planck-Institut f\"{u}r Physik, Munich, Germany\\
$^{\rm III}$ Also at: Italian National Agency for New Technologies, Energy and Sustainable Economic Development (ENEA), Bologna, Italy\\
$^{\rm IV}$ Also at: Dipartimento DET del Politecnico di Torino, Turin, Italy\\
$^{\rm V}$ Also at: Department of Applied Physics, Aligarh Muslim University, Aligarh, India\\
$^{\rm VI}$ Also at: Institute of Theoretical Physics, University of Wroclaw, Poland\\
$^{\rm VII}$ Also at: An institution covered by a cooperation agreement with CERN\\

\section*{Collaboration Institutes}

$^{1}$ A.I. Alikhanyan National Science Laboratory (Yerevan Physics Institute) Foundation, Yerevan, Armenia\\
$^{2}$ AGH University of Science and Technology, Cracow, Poland\\
$^{3}$ Bogolyubov Institute for Theoretical Physics, National Academy of Sciences of Ukraine, Kiev, Ukraine\\
$^{4}$ Bose Institute, Department of Physics  and Centre for Astroparticle Physics and Space Science (CAPSS), Kolkata, India\\
$^{5}$ California Polytechnic State University, San Luis Obispo, California, United States\\
$^{6}$ Central China Normal University, Wuhan, China\\
$^{7}$ Centro de Aplicaciones Tecnol\'{o}gicas y Desarrollo Nuclear (CEADEN), Havana, Cuba\\
$^{8}$ Centro de Investigaci\'{o}n y de Estudios Avanzados (CINVESTAV), Mexico City and M\'{e}rida, Mexico\\
$^{9}$ Chicago State University, Chicago, Illinois, United States\\
$^{10}$ China Institute of Atomic Energy, Beijing, China\\
$^{11}$ Chungbuk National University, Cheongju, Republic of Korea\\
$^{12}$ Comenius University Bratislava, Faculty of Mathematics, Physics and Informatics, Bratislava, Slovak Republic\\
$^{13}$ COMSATS University Islamabad, Islamabad, Pakistan\\
$^{14}$ Creighton University, Omaha, Nebraska, United States\\
$^{15}$ Department of Physics, Aligarh Muslim University, Aligarh, India\\
$^{16}$ Department of Physics, Pusan National University, Pusan, Republic of Korea\\
$^{17}$ Department of Physics, Sejong University, Seoul, Republic of Korea\\
$^{18}$ Department of Physics, University of California, Berkeley, California, United States\\
$^{19}$ Department of Physics, University of Oslo, Oslo, Norway\\
$^{20}$ Department of Physics and Technology, University of Bergen, Bergen, Norway\\
$^{21}$ Dipartimento di Fisica, Universit\`{a} di Pavia, Pavia, Italy\\
$^{22}$ Dipartimento di Fisica dell'Universit\`{a} and Sezione INFN, Cagliari, Italy\\
$^{23}$ Dipartimento di Fisica dell'Universit\`{a} and Sezione INFN, Trieste, Italy\\
$^{24}$ Dipartimento di Fisica dell'Universit\`{a} and Sezione INFN, Turin, Italy\\
$^{25}$ Dipartimento di Fisica e Astronomia dell'Universit\`{a} and Sezione INFN, Bologna, Italy\\
$^{26}$ Dipartimento di Fisica e Astronomia dell'Universit\`{a} and Sezione INFN, Catania, Italy\\
$^{27}$ Dipartimento di Fisica e Astronomia dell'Universit\`{a} and Sezione INFN, Padova, Italy\\
$^{28}$ Dipartimento di Fisica `E.R.~Caianiello' dell'Universit\`{a} and Gruppo Collegato INFN, Salerno, Italy\\
$^{29}$ Dipartimento DISAT del Politecnico and Sezione INFN, Turin, Italy\\
$^{30}$ Dipartimento di Scienze MIFT, Universit\`{a} di Messina, Messina, Italy\\
$^{31}$ Dipartimento Interateneo di Fisica `M.~Merlin' and Sezione INFN, Bari, Italy\\
$^{32}$ European Organization for Nuclear Research (CERN), Geneva, Switzerland\\
$^{33}$ Faculty of Electrical Engineering, Mechanical Engineering and Naval Architecture, University of Split, Split, Croatia\\
$^{34}$ Faculty of Engineering and Science, Western Norway University of Applied Sciences, Bergen, Norway\\
$^{35}$ Faculty of Nuclear Sciences and Physical Engineering, Czech Technical University in Prague, Prague, Czech Republic\\
$^{36}$ Faculty of Physics, Sofia University, Sofia, Bulgaria\\
$^{37}$ Faculty of Science, P.J.~\v{S}af\'{a}rik University, Ko\v{s}ice, Slovak Republic\\
$^{38}$ Frankfurt Institute for Advanced Studies, Johann Wolfgang Goethe-Universit\"{a}t Frankfurt, Frankfurt, Germany\\
$^{39}$ Fudan University, Shanghai, China\\
$^{40}$ Gangneung-Wonju National University, Gangneung, Republic of Korea\\
$^{41}$ Gauhati University, Department of Physics, Guwahati, India\\
$^{42}$ Helmholtz-Institut f\"{u}r Strahlen- und Kernphysik, Rheinische Friedrich-Wilhelms-Universit\"{a}t Bonn, Bonn, Germany\\
$^{43}$ Helsinki Institute of Physics (HIP), Helsinki, Finland\\
$^{44}$ High Energy Physics Group,  Universidad Aut\'{o}noma de Puebla, Puebla, Mexico\\
$^{45}$ Horia Hulubei National Institute of Physics and Nuclear Engineering, Bucharest, Romania\\
$^{46}$ Indian Institute of Technology Bombay (IIT), Mumbai, India\\
$^{47}$ Indian Institute of Technology Indore, Indore, India\\
$^{48}$ INFN, Laboratori Nazionali di Frascati, Frascati, Italy\\
$^{49}$ INFN, Sezione di Bari, Bari, Italy\\
$^{50}$ INFN, Sezione di Bologna, Bologna, Italy\\
$^{51}$ INFN, Sezione di Cagliari, Cagliari, Italy\\
$^{52}$ INFN, Sezione di Catania, Catania, Italy\\
$^{53}$ INFN, Sezione di Padova, Padova, Italy\\
$^{54}$ INFN, Sezione di Pavia, Pavia, Italy\\
$^{55}$ INFN, Sezione di Torino, Turin, Italy\\
$^{56}$ INFN, Sezione di Trieste, Trieste, Italy\\
$^{57}$ Inha University, Incheon, Republic of Korea\\
$^{58}$ Institute for Gravitational and Subatomic Physics (GRASP), Utrecht University/Nikhef, Utrecht, Netherlands\\
$^{59}$ Institute of Experimental Physics, Slovak Academy of Sciences, Ko\v{s}ice, Slovak Republic\\
$^{60}$ Institute of Physics, Homi Bhabha National Institute, Bhubaneswar, India\\
$^{61}$ Institute of Physics of the Czech Academy of Sciences, Prague, Czech Republic\\
$^{62}$ Institute of Space Science (ISS), Bucharest, Romania\\
$^{63}$ Institut f\"{u}r Kernphysik, Johann Wolfgang Goethe-Universit\"{a}t Frankfurt, Frankfurt, Germany\\
$^{64}$ Instituto de Ciencias Nucleares, Universidad Nacional Aut\'{o}noma de M\'{e}xico, Mexico City, Mexico\\
$^{65}$ Instituto de F\'{i}sica, Universidade Federal do Rio Grande do Sul (UFRGS), Porto Alegre, Brazil\\
$^{66}$ Instituto de F\'{\i}sica, Universidad Nacional Aut\'{o}noma de M\'{e}xico, Mexico City, Mexico\\
$^{67}$ iThemba LABS, National Research Foundation, Somerset West, South Africa\\
$^{68}$ Jeonbuk National University, Jeonju, Republic of Korea\\
$^{69}$ Johann-Wolfgang-Goethe Universit\"{a}t Frankfurt Institut f\"{u}r Informatik, Fachbereich Informatik und Mathematik, Frankfurt, Germany\\
$^{70}$ Korea Institute of Science and Technology Information, Daejeon, Republic of Korea\\
$^{71}$ KTO Karatay University, Konya, Turkey\\
$^{72}$ Laboratoire de Physique des 2 Infinis, Ir\`{e}ne Joliot-Curie, Orsay, France\\
$^{73}$ Laboratoire de Physique Subatomique et de Cosmologie, Universit\'{e} Grenoble-Alpes, CNRS-IN2P3, Grenoble, France\\
$^{74}$ Lawrence Berkeley National Laboratory, Berkeley, California, United States\\
$^{75}$ Lund University Department of Physics, Division of Particle Physics, Lund, Sweden\\
$^{76}$ Nagasaki Institute of Applied Science, Nagasaki, Japan\\
$^{77}$ Nara Women{'}s University (NWU), Nara, Japan\\
$^{78}$ National and Kapodistrian University of Athens, School of Science, Department of Physics , Athens, Greece\\
$^{79}$ National Centre for Nuclear Research, Warsaw, Poland\\
$^{80}$ National Institute of Science Education and Research, Homi Bhabha National Institute, Jatni, India\\
$^{81}$ National Nuclear Research Center, Baku, Azerbaijan\\
$^{82}$ National Research and Innovation Agency - BRIN, Jakarta, Indonesia\\
$^{83}$ Niels Bohr Institute, University of Copenhagen, Copenhagen, Denmark\\
$^{84}$ Nikhef, National institute for subatomic physics, Amsterdam, Netherlands\\
$^{85}$ Nuclear Physics Group, STFC Daresbury Laboratory, Daresbury, United Kingdom\\
$^{86}$ Nuclear Physics Institute of the Czech Academy of Sciences, Husinec-\v{R}e\v{z}, Czech Republic\\
$^{87}$ Oak Ridge National Laboratory, Oak Ridge, Tennessee, United States\\
$^{88}$ Ohio State University, Columbus, Ohio, United States\\
$^{89}$ Physics department, Faculty of science, University of Zagreb, Zagreb, Croatia\\
$^{90}$ Physics Department, Panjab University, Chandigarh, India\\
$^{91}$ Physics Department, University of Jammu, Jammu, India\\
$^{92}$ Physics Department, University of Rajasthan, Jaipur, India\\
$^{93}$ Physics Program and International Institute for Sustainability with Knotted Chiral Meta Matter (SKCM2), Hiroshima University, Hiroshima, Japan\\
$^{94}$ Physikalisches Institut, Eberhard-Karls-Universit\"{a}t T\"{u}bingen, T\"{u}bingen, Germany\\
$^{95}$ Physikalisches Institut, Ruprecht-Karls-Universit\"{a}t Heidelberg, Heidelberg, Germany\\
$^{96}$ Physik Department, Technische Universit\"{a}t M\"{u}nchen, Munich, Germany\\
$^{97}$ Politecnico di Bari and Sezione INFN, Bari, Italy\\
$^{98}$ Research Division and ExtreMe Matter Institute EMMI, GSI Helmholtzzentrum f\"ur Schwerionenforschung GmbH, Darmstadt, Germany\\
$^{99}$ Saha Institute of Nuclear Physics, Homi Bhabha National Institute, Kolkata, India\\
$^{100}$ School of Physics and Astronomy, University of Birmingham, Birmingham, United Kingdom\\
$^{101}$ Secci\'{o}n F\'{\i}sica, Departamento de Ciencias, Pontificia Universidad Cat\'{o}lica del Per\'{u}, Lima, Peru\\
$^{102}$ Stefan Meyer Institut f\"{u}r Subatomare Physik (SMI), Vienna, Austria\\
$^{103}$ SUBATECH, IMT Atlantique, Nantes Universit\'{e}, CNRS-IN2P3, Nantes, France\\
$^{104}$ Suranaree University of Technology, Nakhon Ratchasima, Thailand\\
$^{105}$ Technical University of Ko\v{s}ice, Ko\v{s}ice, Slovak Republic\\
$^{106}$ The Henryk Niewodniczanski Institute of Nuclear Physics, Polish Academy of Sciences, Cracow, Poland\\
$^{107}$ The University of Texas at Austin, Austin, Texas, United States\\
$^{108}$ Universidad Aut\'{o}noma de Sinaloa, Culiac\'{a}n, Mexico\\
$^{109}$ Universidade de S\~{a}o Paulo (USP), S\~{a}o Paulo, Brazil\\
$^{110}$ Universidade Estadual de Campinas (UNICAMP), Campinas, Brazil\\
$^{111}$ Universidade Federal do ABC, Santo Andre, Brazil\\
$^{112}$ University of Cape Town, Cape Town, South Africa\\
$^{113}$ University of Houston, Houston, Texas, United States\\
$^{114}$ University of Jyv\"{a}skyl\"{a}, Jyv\"{a}skyl\"{a}, Finland\\
$^{115}$ University of Kansas, Lawrence, Kansas, United States\\
$^{116}$ University of Liverpool, Liverpool, United Kingdom\\
$^{117}$ University of Science and Technology of China, Hefei, China\\
$^{118}$ University of South-Eastern Norway, Kongsberg, Norway\\
$^{119}$ University of Tennessee, Knoxville, Tennessee, United States\\
$^{120}$ University of the Witwatersrand, Johannesburg, South Africa\\
$^{121}$ University of Tokyo, Tokyo, Japan\\
$^{122}$ University of Tsukuba, Tsukuba, Japan\\
$^{123}$ University Politehnica of Bucharest, Bucharest, Romania\\
$^{124}$ Universit\'{e} Clermont Auvergne, CNRS/IN2P3, LPC, Clermont-Ferrand, France\\
$^{125}$ Universit\'{e} de Lyon, CNRS/IN2P3, Institut de Physique des 2 Infinis de Lyon, Lyon, France\\
$^{126}$ Universit\'{e} de Strasbourg, CNRS, IPHC UMR 7178, F-67000 Strasbourg, France, Strasbourg, France\\
$^{127}$ Universit\'{e} Paris-Saclay Centre d'Etudes de Saclay (CEA), IRFU, D\'{e}partment de Physique Nucl\'{e}aire (DPhN), Saclay, France\\
$^{128}$ Universit\`{a} degli Studi di Foggia, Foggia, Italy\\
$^{129}$ Universit\`{a} del Piemonte Orientale, Vercelli, Italy\\
$^{130}$ Universit\`{a} di Brescia, Brescia, Italy\\
$^{131}$ Variable Energy Cyclotron Centre, Homi Bhabha National Institute, Kolkata, India\\
$^{132}$ Warsaw University of Technology, Warsaw, Poland\\
$^{133}$ Wayne State University, Detroit, Michigan, United States\\
$^{134}$ Westf\"{a}lische Wilhelms-Universit\"{a}t M\"{u}nster, Institut f\"{u}r Kernphysik, M\"{u}nster, Germany\\
$^{135}$ Wigner Research Centre for Physics, Budapest, Hungary\\
$^{136}$ Yale University, New Haven, Connecticut, United States\\
$^{137}$ Yonsei University, Seoul, Republic of Korea\\
$^{138}$  Zentrum  f\"{u}r Technologie und Transfer (ZTT), Worms, Germany\\
$^{139}$ Affiliated with an institute covered by a cooperation agreement with CERN\\
$^{140}$ Affiliated with an international laboratory covered by a cooperation agreement with CERN\\

\end{flushleft} 

\end{document}